\documentclass[prl,tightenlines,showpacs,nofootinbib,amsfonts,amssymb,amsmath,twocolumn,widetext]{revtex4}

\usepackage{graphicx}
\usepackage{amsmath}
\usepackage{amsfonts} 
\usepackage{amssymb}  
\usepackage{mathrsfs}
\usepackage{epsfig,bbm}
\usepackage{bm}
\usepackage{color}
\usepackage{dsfont}
\usepackage{tensor}
\usepackage{xspace}

\usepackage{wasysym}
\usepackage{amsthm}
\usepackage{hyperref}

\begin{document}
\newcommand{\average}[1]{\langle{#1}\rangle_{{\cal D}}}
\newcommand{\dd}{{\rm d}}
\newcommand{\etal}{{\it et al.}\xspace}
\newcommand{\e}[1]{_{\text{#1}}}
\newcommand{\h}[1]{^{\text{#1}}}
\newcommand{\ph}{\varphi}
\newcommand{\eps}{\varepsilon}
\newcommand{\hreftt}[2]{\href{#1}{\texttt{#2}}}

\title{Can all cosmological observations be accurately interpreted with a unique geometry?}

\author{Pierre Fleury$^{1,2}$}
\email{fleury@iap.fr}       
             
\author{H\'el\`ene Dupuy$^{1,2,3}$}
\email{helene.dupuy@cea.fr}         

\author{Jean-Philippe Uzan$^{1,2}$}
\email{uzan@iap.fr}
 
\affiliation{
$^1$ Institut d'Astrophysique de Paris, UMR-7095 du CNRS, Universit\'e Pierre et Marie Curie, 98 bis bd Arago, 75014 	Paris, France.\\
$^2$ Sorbonne Universit\'es, Institut Lagrange de Paris, 98 bis bd Arago, 75014 Paris, France.\\
$^3$ Institut de Physique Th\'eorique,
	 CEA, IPhT, URA 2306 CNRS, F-91191 Gif-sur-Yvette, France.
}

\begin{abstract}
The recent analysis of the Planck results reveals a tension between the best fits for $(\Omega_{\rm m0}, H_0)$ derived from the cosmic microwave background or baryonic acoustic oscillations on the one hand, and the Hubble diagram on the other hand. These observations probe the universe on very different scales since they involve light beams of very different angular sizes, hence the tension between them may indicate that they should not be interpreted the same way. More precisely, this letter questions the accuracy of using only the (perturbed) Friedmann-Lema\^itre geometry to interpret all the cosmological observations, regardless of their angular or spatial resolution. We show that using an inhomogeneous ``Swiss-cheese'' model to interpret the Hubble diagram allows us to reconcile the inferred value of $\Omega\e{m0}$ with the Planck results. Such an approach does not require us to invoke new physics nor to violate the Copernican principle.

\end{abstract}

\date{\today}
\maketitle

The standard interpretation of cosmological data relies on the description of the Universe by a spatially homogeneous and isotropic spacetime with a Friedmann-Lema\^{\i}tre (FL) geometry, allowing for perturbations~\cite{pubook}. The emergence of a dark sector, including dark matter and dark energy, emphasizes the need for extra degrees of freedom, either physical (new fundamental fields or interactions) or geometrical (e.g. a cosmological solution with lower symmetry). This has driven a lot of activity to test the hypotheses~\cite{jputesthyp} of the cosmological model, such as general relativity or the Copernican principle.

The recent Planck data were analyzed in such a framework~\cite{planckXVI} in which the cosmic microwave background (CMB) anisotropies are treated as perturbations around a FL universe, with most of the analysis performed at linear order. Nonlinear effects remain small~\cite{pub} and below the constraints on non-Gaussianity derived by Planck~\cite{planckXXIX}. The results nicely confirm the standard cosmological model of a spatially Euclidean FL universe with a cosmological constant, dark matter and initial perturbations compatible with the predictions of inflation.

Among the constraints derived from the CMB, the Hubble parameter $H_0$ and the matter density parameter $\Omega_{\rm m0}$ are mostly constrained through the combination $\Omega_{\rm m0}h^3$, where $H_0=h\times100~{\rm km/s/Mpc}$. It is set by the acoustic scale $\theta_*=r_{\rm s}/D_{\rm A}$, defined as the ratio between the sound horizon and the angular distance at the time of last scattering. The measurement of seven acoustic peaks enables one to determine $\theta_*$ with a precision better than~0.1\%. The constraints on the plane ($\Omega_{\rm m0}, H_0$) are presented in Fig.~3 of Ref.~\cite{planckXVI} and clearly show this degeneracy. The marginalized constraints on the two parameters were then derived~\cite{planckXVI} to be
\begin{equation}\label{planck.Omegam}
 H_0 =(67.3\pm1.2)~{\rm km/s/Mpc},\quad
 \Omega_{\rm m0}=0.315\pm0.017
\end{equation}
at a 68\% confidence level. It was pointed out (see Secs.~5.2--5.4 of Ref.~\cite{planckXVI}) that the values of $H_0$ and $\Omega_{\rm m0}$ are, respectively, low and high compared with their values inferred from the Hubble diagram. Such a trend was already indicated by WMAP-9~\cite{wmap9} which concluded $H_0 =(70\pm2.2)~{\rm km/s/Mpc}$.

Regarding the Hubble constant, two astrophysical measurements are in remarkable agreement. First, the estimation based on the distance ladder calibrated by three different techniques (masers, Milky Way cepheids, and Large Magellanic Cloud cepheids) gives~\cite{freeman2012} $H_0 =(74.3\pm1.5\pm2.1)~{\rm km/s/Mpc}$, respectively with statistical and systematic errors. This improves the earlier constraint obtained by the HST Key program~\cite{freeman2001}, $H_0 =(72\pm8)~{\rm km/s/Mpc}$. Second, the Hubble diagram of type Ia supernovae (SNe Ia) calibrated with the Hubble Space Telescope (HST) observations of cepheids leads~\cite{riess} to $H_0 =(73.8\pm2.4)~{\rm km/s/Mpc}$. Other determinations of the Hubble constant, e.g., from very-long-baseline interferometry observations~\cite{vlbi} or from the combination of Sunyaev-Zel'dovich effect and X-ray observations~\cite{szh0}, have larger error bars and are compatible with both the CMB and distance measurements.

Additionally, the analsis of the Hubble diagram of SNe~Ia leads to a lower value of $\Omega_{\rm m0}$---e.g. $0.222\pm0.034$ from SNLS 3~\cite{SNLS3}---compared to the constraint~(\ref{planck.Omegam}) by Planck. As concluded in Ref.~\cite{planckXVI}, there is no direct inconsistency,  and it was pointed out that there could be residual systematics not properly accounted for in the SN data. Still, it was stated that ``the tension between CMB-based estimates and the astrophysical measurements of $H_0$ is intriguing and merits further discussion.''

Interestingly, the CMB constraints on ($\Omega_{\rm m0},H_0$) are in excellent agreement with baryon acoustic oscillation (BAO) measurements~\cite{bao}, which allow one to determine the angular distance up to redshifts of order 0.7. The common point between the CMB and BAO measurements is that they involve light beams much larger than those involved in astronomical observations. Indeed, a pixel of Planck's high-resolution CMB maps corresponds to 5 arc min \cite{planckI}, while the typical angular size of a SN is $10^{-7}$~arc sec. This means that \emph{the two kinds of observations probe the universe at very different scales}. Moreover, for both the CMB and BAO measurements the crucial information is encoded in correlations, while SN observations rely on ``1-point measurements'' (we are interested in the luminosity and redshift of each SN, not in the correlations between several SNe). Because of such distinctions one can expect the two classes of cosmological observations to be affected differently by the inhomogeneity of the Universe, through gravitational lensing.

The effect of lensing on CMB measurements is essentially due to the large-scale structure, and it can be taken into account in the framework of cosmological perturbation theory at linear order~\cite{planckXVII} (see, however, Ref.~\cite{BolejkoCMB} for a discussion about the impact of strong inhomogeneities). We refer to Ref.~\cite{Vallinotto2007} for a description of the lensing effects on BAO measurements. Regarding the Hubble diagram, the influence of lensing has also been widely investigated~\cite{cemuu}. The propagation of light in an inhomogeneous universe gives rise to both distortion and magnification. Most images are expected to be demagnified because their lines of sight probe underdense regions, while some are amplified due to strong lensing.
It shall thus induce a dispersion of the luminosities of the sources, that is, an extra scatter in the Hubble diagram~\cite{scatter}. Its amplitude can be determined from the perturbation theory~\cite{ptheory} and subtracted~\cite{correction}. However, a considerable fraction of the lensing effects arises from sub-arc-min scales, which are not probed by shear maps smoothed on arc min scales~\cite{smallscale}.

The tension on ($\Omega_{\rm m0},H_0$) may indicate that, given the accuracy of the observations achieved today, the use of a (perturbed) FL geometry to interpret the astrophysical data is no longer adapted. More precisely, the question that we want to raise is {\em whether the use of a unique spacetime geometry is relevant for interpreting all the cosmological observations, regardless of their angular or spatial resolution and of their location (redshift)}. Indeed, each observation is expected to probe the Universe smoothed on a typical scale related to its resolution, and this can lead to fundamentally different geometrical situations. In a universe with a discrete distribution of matter, the Riemann curvature experienced by a beam of test particles or photons is dominated by the Weyl tensor. Conversely, in a (statistically spatially isotropic) universe smoothed on large scales it is dominated by the Ricci tensor. Both situations correspond to distinct optical properties~\cite{ricciweyl}.

In the framework of geometric optics, a light beam is described by  a bundle of null geodesics. All the information about the size and the shape of a beam can be encoded in a $2 \times 2$ matrix ${\cal D}\indices{^A_B}$ called the Jacobi map (see Ref.~\cite{sw-us} for further details). In particular, the angular and luminosity distances read, respectively,
\begin{equation}\label{eq:distances}
D\e{A} = \sqrt{\vert{\rm det} {\cal D}\indices{^A_B}\vert},
\qquad
D\e{L} = (1+z)^2 D\e{A}.
\end{equation}
where $z$ denotes the redshift. The evolution of the Jacobi map with light propagation is governed by the Sachs equation \cite{sachs,refbooks}
\begin{equation}\label{eq:jacobi}
\frac{\dd^2}{\dd v^2}\,{\cal D}\indices{^A_B}
={\cal R}\indices{^A_C}\,{\cal D}\indices{^C_B} ,
\end{equation}
%
%
where $v$ is an affine parameter along the geodesic bundle. The term ${\cal R}_{AB}$ which controls the evolution of ${\cal D}\indices{^A_B}$ is a projection of the Riemann tensor called the optical tidal matrix. It is defined by ${\cal R}_{AB} \equiv {R}_{\mu\nu\alpha\beta} s_A^\mu k^\nu k^\alpha s_B^\beta$, where $k^\mu$ is the wave vector of an arbitrary ray, and the Sachs basis $\lbrace s_A^\mu\rbrace_{A=1,2}$ spans a screen on which the observer projects the light beam. Because the Riemann tensor can be split into a Ricci part~$R_{\mu\nu}$ and a Weyl part~$C_{\mu\nu\alpha\beta}$, the optical tidal matrix can also be decomposed as
\begin{equation}\label{eq:decomposition_optical_matrix}
(\mathcal{R}_{AB}) =
\underset{\text{Ricci lensing}}{
\underbrace{
\begin{pmatrix}
\Phi_{00} & 0 \\ 0 & \Phi_{00}
\end{pmatrix}
}}
+
\underset{\text{Weyl lensing}}{
\underbrace{
\begin{pmatrix}
- {\rm Re}\,\Psi_0 & {\rm Im}\,\Psi_0 \\
  {\rm Im}\,\Psi_0 & {\rm Re}\,\Psi_0
\end{pmatrix}
}}
\end{equation}
with $\Phi_{00} \equiv -(1/2) R_{\mu\nu} k^\mu k^\nu$ and $\Psi_0 \equiv - (1/2) C_{\mu\nu\alpha\beta} (s_1^{\mu}-{\rm i}s_2^{\mu}) k^\nu k^\alpha (s_1^{\beta}-{\rm i}s_2^{\beta})$. It clearly appears in Eq.~ \eqref{eq:decomposition_optical_matrix} that the Ricci term tends to isotropically focus the light beam, while the Weyl term tends to shear and rotate it. The behavior of a light beam is thus different whether it experiences Ricci-dominated lensing (large beams, e.g. CMB measurements) or Weyl-dominated lensing (narrow beams, e.g. SN observations).

This Ricci-Weyl problem can be addressed with different methods. One possibility, a representative of which is the Dyer-Roeder approximation~\cite{DR}, is to construct a general distance-redshift relation which would take into account the effect of inhomogeneities in some average way. However, such approaches are in general difficult to control~\cite{cemuu} because they rely on approximations whose domain of applicability is unknown. An alternative possibility consists in constructing inhomogeneous cosmological models, with a discrete distribution of matter, and studying the impact on light propagation. Several models exist in the literature: the Schwarzschild-cell method~\cite{archipel} or the lattice universe~\cite{pointd} which are both approximate solutions of the Einstein equations; and the Swiss-cheese models~\cite{sw-original} which are constructed by matching together patches of exact solutions of the Einstein equations. This last approach is the one that we shall follow in this Letter.

Consider a Swiss-cheese model in which clumps of matter (modeling e.g. galaxies), each of them lying at the center of a spherical void, are embedded in a FL spacetime. The interior region of a void is described by the Kottler geometry---i.e., Schwarzschild with a cosmological constant---while the exterior geometry is the FL one. By construction, such inhomogeneities do not modify the expansion dynamics of the embedding FL universe, thus avoiding any discussion regarding backreaction. The resulting spacetime is well defined, because the Darmois-Israel junction conditions are satisfied on the boundary of every void.
Compared to a strictly homogeneous universe, a Swiss-cheese model is therefore characterized by two additional parameters: the size of the voids (or equivalently the mass of their central bodies), and the volumic fraction of the remaining FL regions which encodes the smoothness of the distribution of matter. It is naturally quantified by the \emph{smoothness parameter}
\begin{equation}\label{eq:smoothness_parameter}
f \equiv
\underset{V\rightarrow\infty}{\rm lim} \frac{V\e{FL}}{V},
\end{equation}
where $V\e{FL}$ is the volume occupied by the FL region within a volume $V$ of the Swiss cheese. With the definition \eqref{eq:smoothness_parameter}, $f=1$ corresponds to a model with no hole (i.e. a FL universe), while $f=0$ corresponds to the case where matter is exclusively under the form of clumps inside voids.

Of course such a model cannot be considered realistic, but neither does the exact FL geometry, used to interpret the Hubble diagram. Both spacetimes describe a spatially statistically homogeneous and isotropic universe, and the former permits additionally the investigation of the effect of a discrete distribution of matter. Since the FL universe is a particular Swiss-cheese model, this family of spacetimes therefore allows us to estimate how good the hypothesis of strict spatial homogeneity---with a continuous matter distribution at \emph{all} scales---is.

The propagation of light in a Swiss-cheese model has been comprehensively investigated in Ref.~\cite{sw-us}, generalizing earlier works~\cite{swearly}, with the key assumption that \emph{light never crosses the clumps}. This ``opacity assumption'' can be observationally justified in the case of SN observations if the clumps represent galaxies (see Ref.~\cite{sw-us} for a discussion). Compared to the strictly homogeneous case, any light signal traveling through a Swiss cheese then experiences a reduced  Ricci focusing. This leads [see Eqs.~\eqref{eq:distances}-\eqref{eq:decomposition_optical_matrix}] to an increase of the observed luminosity distance~$D\e{L}$. The effect of Weyl lensing---i.e. here shear---is relatively small.

This systematic effect, due to inhomogeneities, tends to bias the Hubble diagram in a way that mimics the contribution of a negative spatial curvature or a positive cosmological constant. In other words, if one interprets the Hubble diagram of a Swiss-cheese universe by wrongly assuming that it is strictly homogeneous, then one underestimates the value of $\Omega\e{m0}$. The error reaches a few percent, which is comparable to other estimates in similar contexts~\cite{bolejko}. Note, however, that in the case of Swiss-cheese models with Lema\^{\i}tre-Tolman-Bondi patches instead of Kottler voids, the effect of inhomogeneities has a much smaller impact on the Hubble diagram~\cite{sw-ltb}. Thus, the systematic effect exhibited in Ref.~\cite{sw-us} must be attributed to the discreteness of the distribution of matter.

Simulating the mock Hubble diagrams for Swiss-cheese universes with various values of its parameters, we inferred a phenomenological expression for the luminosity distance $D_{\rm L}(z;\Omega_{\rm m0},\Omega_{\Lambda0},H_0,f)$ which is very close to the Dyer-Roeder one. This expression was then used to fit the Hubble diagram constructed from the SNLS~3 catalog~\cite{SNLS3}. Figure~25 of Ref.~\cite{sw-us} shows that $f$ influences the result of the best fit on $\Omega_{\rm m0}$ that can shift from $0.22$ for $f=1$ (in agreement with the standard FL analysis performed in Ref.~\cite{SNLS3}) to $0.3$ for $f=0$.

\begin{figure}[!h]
\includegraphics[width=\columnwidth]{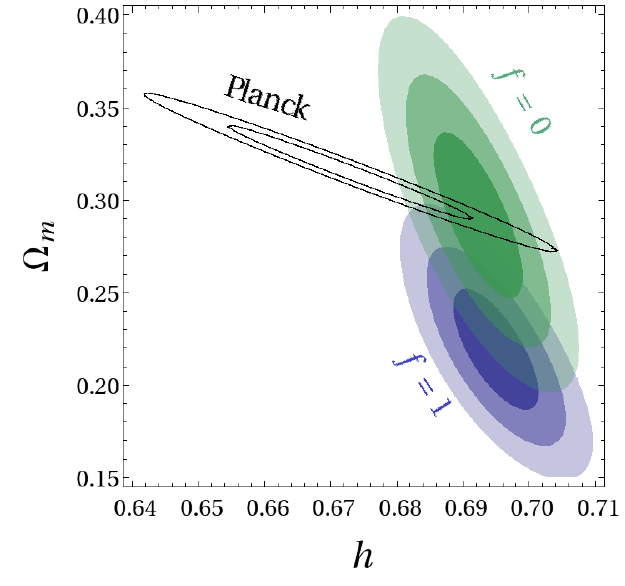}
\caption{Comparison of the constraints obtained by Planck on $(\Omega_{\rm m0}, h)$~\cite{planckXVI} and from the analysis of the Hubble diagram constructed from the SNLS 3 catalog \cite{SNLS3}. The shaded contour plots correspond to two different smoothness parameters. For $f=1$, the geometry used to fit the data is the FL one.}
\label{fig:constraints}
\end{figure}

Figure~\ref{fig:constraints} shows the constraints in the plane $(h,\Omega\e{m0})$ imposed by Planck on the one hand, and by the Hubble diagram on the other hand, whether it is interpreted in a spatially flat FL universe ($f=1$) or in a spatially flat Swiss-cheese model for which matter is entirely clumped ($f=0$). The agreement between the CMB and the Hubble diagram is clearly improved for small values of $f$, especially regarding $\Omega\e{m0}$, while $h$ is almost unaffected.

Note that SN observations \emph{alone} cannot constrain $H_0$, because of the degeneracy with the (unknown) absolute magnitude~$M$ of the SNe. For the results of Fig.~\ref{fig:constraints} the degeneracy was broken by fixing $M=-19.21$, according to the best-fit value obtained by Ref.~\cite{SNLS3} with a fiducial Hubble constant $h=0.7$. Thus the horizontal positions of the SN contours in Fig.~\ref{fig:constraints} are only \emph{indicative}.

Alleviating the tension on $H_0$ remains an open issue. Because inferring its value from SNe is a \emph{local} measurement, a promising approach consists in taking into account the impact of our close environment. It has been suggested \cite{Marra13} that cosmic variance increases the uncertainty on $H_0\h{local}$ and thus reduces the tension with $H_0\h{CMB}$. More speculatively, $H_0\h{local}>H_0\h{CMB}$ may be a hint that our local environment is underdense \cite{Jha2007}. Our conclusions on $\Omega\e{m0}$ remains, however, unaffected by this issue.

Our analysis, though relying on a particular class of models, indicates that the FL geometry is probably too simplistic to describe the Universe for certain types of observations, given the accuracy reached today.
In the end, a single metric may not be sufficient to describe all the cosmological observations, just as Lilliputians and Brobdingnag's giants \cite{gulliver} cannot use a map with the same resolution to travel. A better cosmological model probably requires an atlas of maps with various smoothing scales, determined by the observations at hand.

Other observations, such as lensing~\cite{voidslensing}, may help to characterize the distribution and the geometry of voids~\cite{voidsben}, in order to construct a better geometrical model. For the first time, the standard FL background geometry may be showing its limits to interpret the cosmological data with the accuracy they require.\\

We thank Francis Bernardeau, Yannick Mellier, Alice Pisani, Cyril Pitrou, Joe Silk and Benjamin Wandelt for discussions. This work was supported by French state funds managed by the ANR within the Investissements d'Avenir programme under reference ANR-11-IDEX-0004-02, the Programme National Cosmologie et Galaxies, and the ANR THALES (ANR-10-BLAN-0507-01-02).

\end{document}